\documentclass{ceab}    
\usepackage{graphicx}   
\usepackage{ceabbib}    
\usepackage[T1]{fontenc}
\usepackage{exscale}    
\usepackage{amstext}    
\usepackage{url}        
\usepackage{units}      
\usepackage[tight]{subfigure}  
\begin{document}
 \typeout{}
 \typeout{********************************************************}
 \typeout{Document information}
 \typeout{Authors:}
 \typeout{         Hakan \"O{}nel (honel@aip.de)}
 \typeout{         Gottfied J. Mann (gmann@aip.de)}
 \typeout{--------------------------------------------------------}
 \typeout{Date of first submission: November 13, 2008}
 \typeout{Date of second submission: February 26, 2009}
 \typeout{--------------------------------------------------------}
 \typeout{Published in:}
 \typeout{Cent. Eur. Astropys. Bull. 33 (2009), 1, 141-154.}
 \typeout{********************************************************}
 \typeout{ Begin of document                                      }
 \typeout{********************************************************}
 \typeout{}
 \newcommand{\mytext}[1]{\ensuremath{\text{#1}}} 
 \newcommand{\dd}[1]{\mytext{d}#1} 
 \newcommand{\ddfrac}[2]{\frac{\mytext{d}#1}{\mytext{d}#2}} 
 \newcommand{\ddnicefrac}[2]{\nicefrac{\mytext{d}#1}{\mytext{d}#2}} 
 \newcommand{\myvec}[1]{\ensuremath{\vec{#1}}} 
 \newcommand{\therm}[0]{\ensuremath{\mytext{therm}}} 
 \mathchardef\ordinarycolon\mathcode`\:
 \mathcode`\:=\string"8000
 \begingroup \catcode`\:=\active
   \gdef:{\mathrel{\mathop\ordinarycolon}}
 \endgroup
 \title{Generation of large scale electric fields in coronal flare circuits}
 \renewcommand{\gore}{Large scale electric fields in coronal circuits}
 \author{H.~\"O{}nel$^\dagger$ and G.~J.~Mann$^*$
         \vspace{2mm}\\
         \it Astrophysical Institute Potsdam\\
         \it An der Sternwarte 16, 14482 Potsdam, Federal Republic of Germany\\
         \it $^\dagger$eMail:~\url{honel@aip.de} -- $^*$eMail:~\url{gmann@aip.de}
	 \vspace*{2mm}\\ Published in: {\bf Cent. Eur. Astropys. Bull. 33 (2009), 1, 141-154.} 
 }
\maketitle
 \begin{abstract}
   A large number of energetic electrons are generated during solar flares. They 
   carry a substantial part of the flare released energy but how these electrons 
   are created is not fully understood yet.

   This paper suggests that plasma motion in an active 
   region in the photosphere is the source of large electric currents. These currents 
   can be described by macroscopic circuits.
   Under special circumstances currents can establish in the corona
   along magnetic field lines.
   The energy released by these currents when moderate assumptions for
   the local conditions are made, is found be comparable to the flare energy.
 \end{abstract}
%
%
 \typeout{=====================================================}
 \typeout{Introduction}
 \section{Introduction}\label{sec_introduction}
   \begin{figure}[t]
     \begin{center}
       \includegraphics[width=0.6\textwidth]{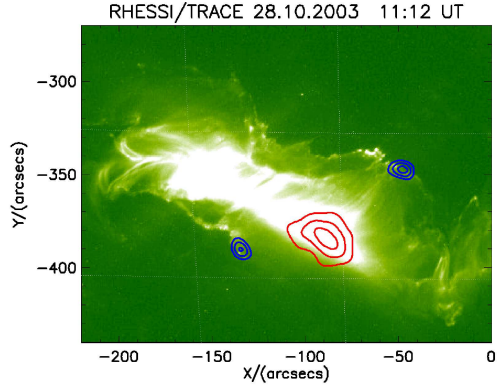}
       \caption[\textit{RHESSI} observations from October~28,~2003.]{The figure shows a flare ribbon as observed by \textit{TRACE} spacecraft on October~28,~2003. The contour plots are obtained by \textit{RHESSI}, \mbox{i.e.}, the loop-top soft \mbox{X-ray} source is located in the middle and the footpoint hard \mbox{X-ray} sources are located beside the \mbox{loop-top} source.}
       \label{img_rhessi}
     \end{center}
   \end{figure}
   \begin{figure}[t]
     \begin{center}
       \includegraphics[width=0.6\textwidth]{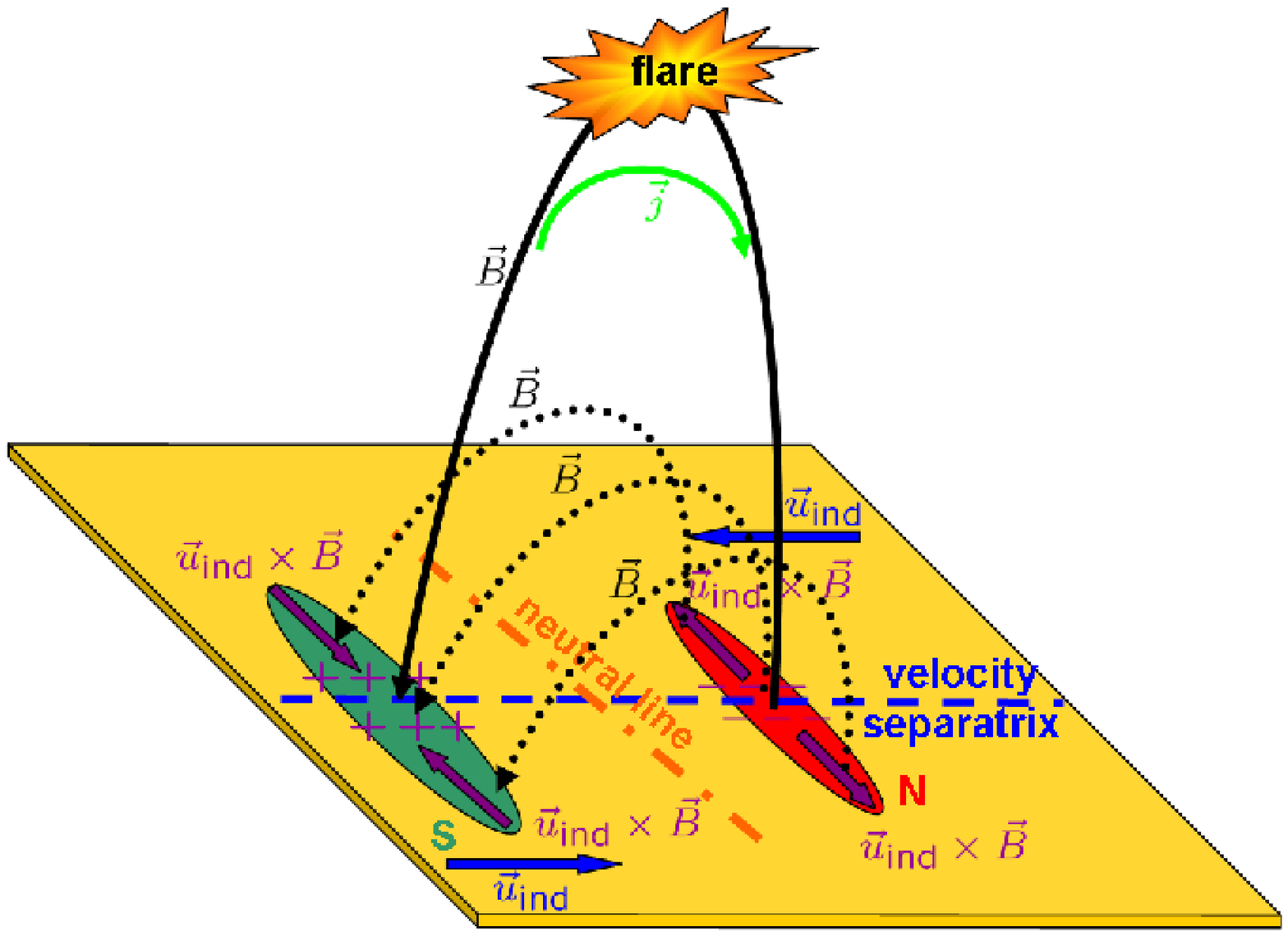}
       \caption[Current Model]{Simplified sketch of the geometrical configuration right even at the flare ignition in the solar corona. The current density \mbox{$\myvec{j}$} is established in the corona as a result of charge separation in the photosphere.}
       \label{img_model}
      \end{center}
    \end{figure}
   In the course of solar flares a large amount of energy is suddenly released
   and transferred into local heating of the coronal plasma, mass motions
   (\mbox{e.g.}, jets and coronal mass ejections), enhanced emission of both
   electromagnetic radiation (from the \mbox{radio-} up to the \mbox{$\gamma$-ray} 
   range)
   and energetic particles (\mbox{i.e.}, electrons, protons, and heavy ions).
   Energetic electrons are
   responsible for the \mbox{non-thermal} radio and \mbox{X-ray}
   emission of the Sun.
   Moreover they carry
   a substantial part of the released flare energy 
   \citep{1971SoPh...17..412L,1974SSRv...16..189L,2004JGRA..10910104E}.
   Oberservations by 
   \textit{RHESSI}\footnote{\textit{RHESSI} stands for the Ramaty High Energy Solar Spectroscopic Imager} 
   indicate that
   flare produced high energetic (\mbox{$\geq \unit[20]{keV}$})
   electron fluxes \mbox{$F_\mytext{e}$} are of the order of 
   \mbox{$F_\mytext{e}\approx\unitfrac[10^{36}]{1}{s}$}
   related to a power \mbox{$P_\mytext{e}$} of about
   \mbox{$P_\mytext{e}\approx\unit[10^{22}]{W}=\unitfrac[10^{29}]{erg}{s}$}
   \citep[][]{1963QB528.S6.......,1974SoPh...38..419H,1974SSRv...16..189L,2004JGRA..10910104E,2007CEAB...31..135W}.
   However it is still unclear how so many electrons are accelerated up to high
   energies within fractions of a second.
   
   In this paper a realistic flare scenario is theoretically modelled. It
   considers genuine 
   parameters basing on average observations of flaring active regions,
   \mbox{e.g.}, see the event on October~28,~2003 as presented
   by the \textit{TRACE}\footnote{\textit{TRACE} stands for the Transition Region and Coronal Explorer}
   image in Fig.~\ref{img_rhessi}. 
   This image also contains the \textit{RHESSI} \mbox{X-ray}
   contour plots, \mbox{i.e.}, the \mbox{loop-top} soft
   \mbox{X-ray} source is located in the middle of the picture, whereas the 
   footpoint hard \mbox{X-ray} sources are located next to 
   the \mbox{loop-top} source. The following conclusions can be drawn
   from this figure:
   The hard \mbox{X-ray} sources (Fig.~\ref{img_rhessi})
   have a diameter of about \mbox{$\oslash_\mytext{s} = \unit[10 \times 10^{6}]{m}$} 
   resulting in a source area \mbox{$A_\mytext{s} = \pi \left(\nicefrac{\oslash_\mytext{s}}{2}\right)^2 = \unit[78.5 \times 10^{12}]{m^2}$}. 
   Both hard \mbox{X-ray} sources are separated from one another by about
   \mbox{$L_\mytext{s}=\unit[70]{Mm}$} (Fig.~\ref{img_rhessi}).
   If it is assumed that both hard \mbox{X-ray} footpoints are located at the same
   height above the photosphere and belong to one circular magnetic field line, 
   then this magnetic loop has an arc length of
   \mbox{$L_\mytext{co} = \nicefrac{\left(\pi L_\mytext{s}\right)}{2} \approx \unit[110]{Mm}$}.

   The average kinetic energy \mbox{$\overline{W}$} of one energetic electron 
   can be estimated by
     \mbox{$\overline{W}\approx\nicefrac{P_\mytext{e}}{F_\mytext{e}}=\nicefrac{\unit[10^{22}]{W}}{\left(\unit[10^{36}]{s^{-1}}\right)}\approx\unit[62.4]{keV}$}
   corresponding to an average velocity 
   \mbox{$\overline{V}=0.454\,c\approx\unitfrac[136]{Mm}{s}$}, where
   $c$ represents the speed of light.
   This velocity can be used to retrieve the electron density 
   \mbox{$N_\mytext{acc}$} of the accelerated electrons 
   \mbox{$N_\mytext{acc}  \approx  \nicefrac{ F_\mytext{e} }{ \left(2 A_\mytext{s}\overline{V}\right) } = \unit[1.17\times 10^{13}]{m^{-3}}$}.  
   The $2$ in this last equation originates 
   from the fact that \textit{RHESSI} usually 
   observes two hard \mbox{X-ray} sources at the 
   footpoints (as it can be seen in Fig.~\ref{img_rhessi}). 
   By assuming a typical electron density 
   \mbox{$N_\mytext{co}=\unit[10^{15}]{m^{-3}}$} in the flare 
   region \citep[see e.g.,][]{2002SSRv..101....1A},
   \mbox{i.e.}, considering the density corresponding to the electron plasma frequency
   of about $\unit[300]{MHz}$,
   it can be seen that only a fraction of the available electrons is finally
   accelerated, \mbox{i.e.}, 
     \mbox{$N_\mytext{acc} \approx 1.2\% \, N_\mytext{co}$}.
   However the energy contained in the accelerated electrons
   in comparison to the \mbox{plasma} \mbox{electrons'} thermal energy 
   (three degrees of freedom are assumed)
   in the
   flare region can be determined by
   \mbox{$\nicefrac{\left(N_\mytext{acc} \overline{W}\right)}{\left(1.5 N_\mytext{co} k_\mytext{B} T\right)} \approx 14.1\%{}$}.
   The quantity \mbox{$k_\mytext{B}$} stands for \mbox{Boltzmann's} constant.
   Here a typical flare temperature \mbox{$T = \unit[40]{MK}$} has been adopted,
   which is a value obtained from the photon fluxes observed by \textit{RHESSI} 
   \citep{2007CEAB...31..135W}. Of course these values represent only
   rough estimates, but they impressively demonstrate that during flares 
   just a small number of electrons are accelerated to 
   relatively high energies.

   Currently several different electron acceleration mechanisms in the 
   solar corona
   are known. All of these mechanisms have the principle of acceleration
   due to electric fields in common, but differ in the processes
   leading to the generation of the electric field.
   In the present paper the generation of a large scale DC~electric field 
   is discussed in terms of electric circuits, which is related to 
   a current generated due to photospheric plasma motion 
   \citep[e.g.,][]{1967SoPh....1..220A,1972SoPh...23..146S,1973SoPh...32..365M,1974SoPh...38..419H,1975SoPh...40..217O,1979SoPh...64..333A,1983SoPh...84..153K,1997ApJ...486..521M,1998A&A...337..887Z,2004ApJ...617L.151Y,2005AstL...31..620Z}.
   Motivated by these papers, the electric currents 
   are investigated in order to obtain a mechanism for 
   the acceleration of electrons to high energies. 
   The basic idea of this mechanism is to generate the 
   flare energy by photospheric plasma motion in active regions.
   This is in contradiction to the reconnection model
   in which the magnetic field energy in the corona 
   is taken for the flare.

 \typeout{=====================================================}
 \typeout{Description of the Model}
 \section{Description of the Model}\label{sec_phenomenology}
  \begin{figure}[t]
     \begin{center}
       \hfill
       \subfigure[Translation of Fig.~\ref{img_model} into a circuit diagram.]{\label{img_complexcircuit}\includegraphics[width=0.45\textwidth]{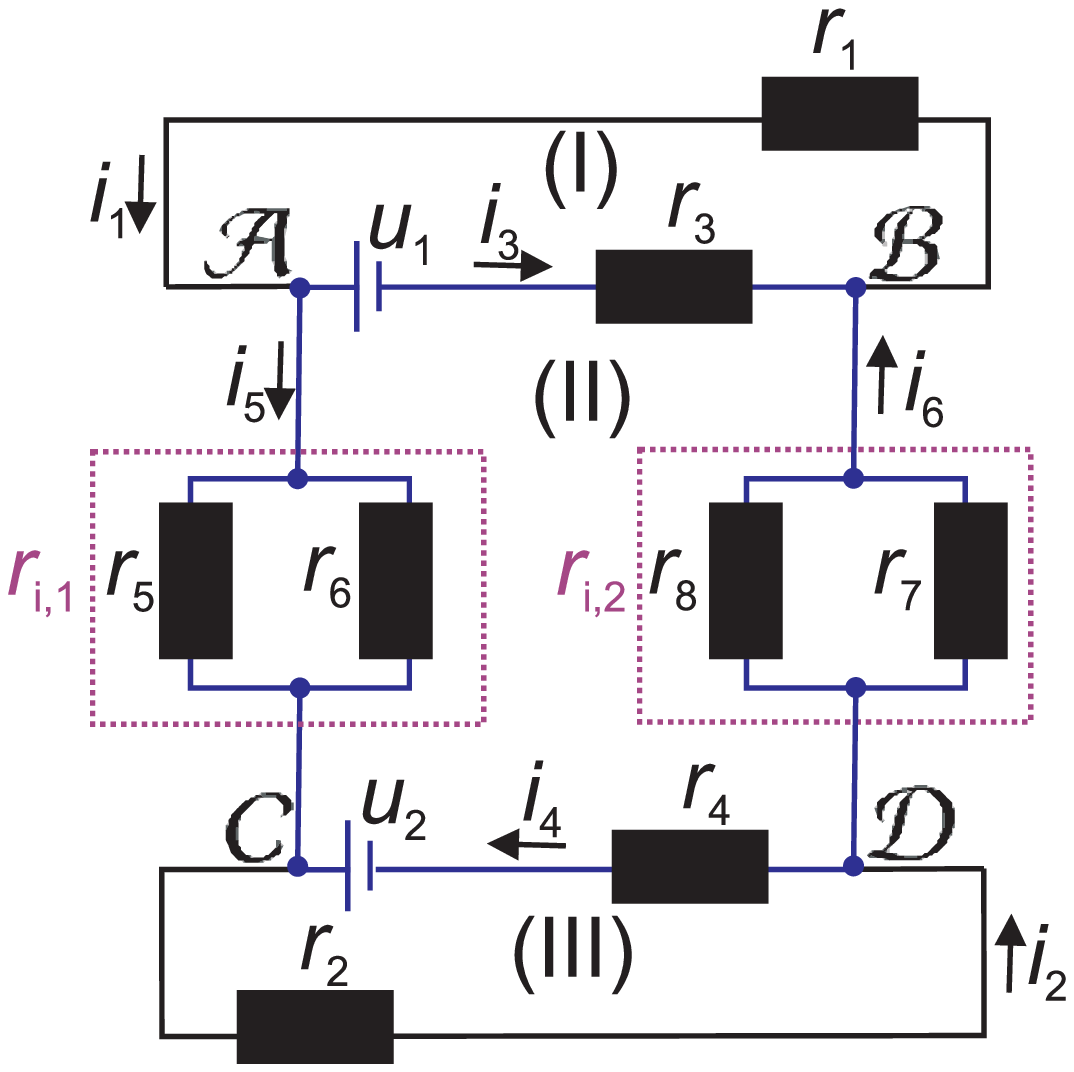}}
       \hfill
       \subfigure[The simplified electric circuit is extracted from the (II)nd bluely coloured mesh of the circuit in Fig.~\ref{img_complexcircuit}.]{\label{img_circuit_secondmesh}\includegraphics[width=0.45\textwidth]{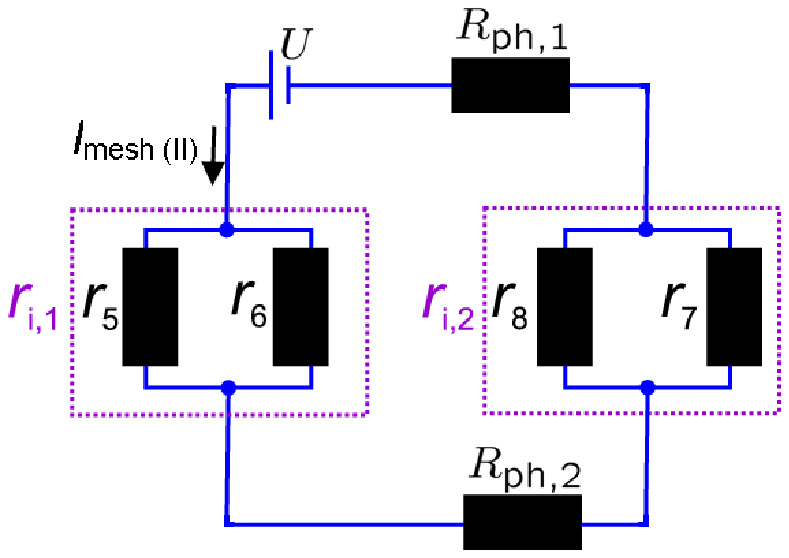}}
       \hfill
       \caption[Circuit diagrams]{Electric circuit diagrams}
       \label{img_thecircuitdiagrams}
      \end{center}
    \end{figure}
   Consider a bipolar active region in the photosphere. The magnetic loops 
   built arc like structures connecting regions of different magnetic polarities. 
   These regions are separated from each other by the magnetic neutral line.
   Since the temperature in the photosphere is about \mbox{$\unit[5.8]{kK}$}
   the plasma is only partly ionised, whereas in the corona
   the temperature is sufficiently
   high to ionise most of the elements completely.
   The active regions are related to high magnetic field 
   concentration penetrating the
   photosphere. For instance 
   \cite{2004ApJ...617L.151Y} have reported magnetic flux densities
   in the case of NOAA~AR~10486 (October~29,~2003) reaching up to
   \mbox{$\unit[0.15]{T}$}. They also mention photospheric flow motion
   with velocities up to \mbox{$\unitfrac[1.6]{km}{s}$}. 
   These observations are not unusual 
   \citep[see e.g.,][]{1990SoPh..125..321K,2003ApJ...593..564W,2004ApJ...605..931W}.
   Hence it is known that the photospheric matter is in motion
   (\mbox{e.g.}, the Evershed motion \citep{1909MNRAS..69..454E,1968SoPh....4..168C},
     cyclonic (also called vortex like) plasma motion \citep[see e.g.,][]{1973SoPh...32..365M,1992HvaOB..16...29H})
     and often high plasma shear velocities can be seen in the photosphere 
   \citep{2003SoPh..215..261S,2004ApJ...617L.151Y,2004ApJ...607L.131X}.

   In Fig.~\ref{img_model} a bipolar active region is schematically
   presented. The plane represents an area on the solar
   photosphere, where a bipolar active region is located.
   The black dotted arrows represent the magnetic flux density
   \mbox{$\myvec{B}$} connecting the regions of different magnetic
   polarities which are separated by the magnetic neutral line.
   Due to the photospheric plasma motion the Lorentz force 
   \mbox{$q \, \myvec{u}_\mytext{ind} \times \myvec{B}$}
   acts on the charges $q$ of the plasma and leads to generation
   of an electric current.
   If the direction of the plasma velocity 
   \mbox{$\myvec{u}_\mytext{ind}$}
   is reversed at the velocity separatrix (dashed line)
   in Fig.~\ref{img_model}, as it is indicated by the velocity arrows \mbox{$\myvec{u}_\mytext{ind}$},
   then the Lorentz force points
   either toward the velocity separatrix or away from it,
   depending on the directions of the magnetic flux vector
   and the plasma velocity vector.

   The generated current leads to electric circuits:
   Electric 
   currents always choose the path of the lowest resistance. 
   Since the 
   plasma resistivity is highly dependent on the temperature, the 
   conductivity in the corona is about \mbox{$1\,090$} times 
   higher than in the photospheric plasma \citep{1965pfig.book.....S}.
   In such a plasma the charged particles propagate along the 
   magnetic field lines corresponding to electric wires. If 
   there is a magnetic connection between two oppositely charged
   areas through the corona, possibly as a result of magnetic reconnection,
   an electric current can
   close
   the electric circuit
   \citep{1967SoPh....1..220A,1974SoPh...38..419H} through the corona (Fig.~\ref{img_model}).
   Then an electric field occurs along the coronal magnetic field lines
   and 
   acts on the electrons within the coronal loop, and accelerates them
   along the magnetic field up to high energies beyond 
   $\unit[100]{keV}$.

 \typeout{=====================================================}
 \typeout{The Electric Circuit Model}
 \section{The Electric Circuit Model}\label{sec_complexcircuit}
  The model (see Fig.~\ref{img_model})
  described in the previous section is translated into a system of electric
  circuits as drawn in Fig.~\ref{img_complexcircuit}.
  There are two electric power supplies \mbox{$u_{1}$} 
  and \mbox{$u_{2}$} 
  for the both different regions of magnetic polarity at the
  bipolar active region. Each of them has its own internal resistor, 
  namely \mbox{$r_{3}$} and \mbox{$r_{4}$}.
  The induced current can be closed 
  via the photosphere of each region, \mbox{i.e.}, via the 
  resistors \mbox{$r_{1}$} and
  \mbox{$r_{2}$}, \mbox{and/or} by an interconnection between these both 
  regions,
  \mbox{i.e.}, via the resistors 
  \mbox{$r_{\mytext{i},1}$} and \mbox{$r_{\mytext{i},2}$}. These 
  interconnections
  can be established by both through the photosphere, \mbox{i.e.}, via the 
  resistors \mbox{$r_{6}$} and \mbox{$r_{8}$},
  and through the corona via the resistors \mbox{$r_{5}$} and \mbox{$r_{7}$}.
  The latter can only happen, if there is a magnetic connection present 
  between both differently polarised regions of the active
  region through the corona.
  For simplicity, the resistors \mbox{$r_{5}$} and \mbox{$r_{6}$}, 
  as well as \mbox{$r_{7}$} and \mbox{$r_{8}$} are combined to 
  \mbox{$r_{\mytext{i},1} = \nicefrac{r_{5} r_{6}}{(r_{5}+r_{6})}$} and
  \mbox{$r_{\mytext{i},2} = \nicefrac{r_{7} r_{8}}{(r_{7}+r_{8})}$}.
  Note that the resistors $r_{5}$ and $r_{7}$ are coronal
  resistors, whereas the other ones are located in the photosphere. Since
  the resistivity is much lower in the corona than in the photosphere,
  the relationship \mbox{$r_{n} \ll r_{o}$} for all
  \mbox{$n \in\left\{5,7\right\}$} and 
  all \mbox{$o \in\left\{1,2,3,4,6,8\right\}$} is satisfied.
  Therefore \mbox{$r_{\mytext{i},1}$} and \mbox{$r_{\mytext{i},2}$}
  become either \mbox{$r_{\mytext{i},1} \approx r_{5}$}
  or \mbox{$r_{\mytext{i},1} \approx r_{6}$} 
  and \mbox{$r_{\mytext{i},2} \approx r_{7}$} or 
  \mbox{$r_{\mytext{i},2} \approx r_{8}$} 
  depending on whether there is a magnetic connection through the 
  corona or not,
  respectively.

  Applying \mbox{Kirchhoff's} laws
  on the circuit presented in Fig.~\ref{img_complexcircuit}
  the important result 
  \mbox{$i_{5} = i_{6}$}
  is found, 
  \mbox{i.e.}, there are always two equal but oppositely directed
  currents connecting the 
  circuits of both magnetic regions.
  An extended evaluation of the circuit system (Fig.~\ref{img_complexcircuit}) provides the value for 
  the current
  \begin{eqnarray}
     i_{5} \!\!\!&=& \!\!\!\frac{r_{1}(r_{2}+r_{4})u_{1} - r_{2}(r_{1}+r_{3})u_{2}}{     r_{1} r_{3} \left(r_{2} + r_{4}\right) + r_{2} r_{4} \left(r_{1} + r_{3}\right) + \left(r_{\mytext{i},1} + r_{\mytext{i},2}\right) \left(r_{1}+r_{3}\right) \left(r_{2} + r_{4}\right)       } \label{eqn_i5}
  \end{eqnarray}
  interconnecting both regions of the active region \citep{2008PhDT.........9O}.
  According to Eq.~(\ref{eqn_i5}) a fully symmetrical circuit, 
  \mbox{i.e.},
  \mbox{$u_{1} = u_{2}$}, 
  \mbox{$r_{1} = r_{2}$}, and 
  \mbox{$r_{3} = r_{4}$} would directly lead to \mbox{$i_{5} = i_{6} = 0$}.
  This means the electrical circuit would
  be completely closed through the photosphere and no current would 
  flow through
  the interconnecting resistors, neither through the 
  coronal part (\mbox{$r_5$} and \mbox{$r_7$}),
  nor through
  the photospheric part (\mbox{$r_6$} and \mbox{$r_8$}).
  However a minor asymmetry (\mbox{e.g.}, caused by different plasma flow
  velocities in the photosphere \mbox{$u_1 \neq u_2$} and/or different values
  of the resistors $r_{1}$, $r_{2}$, $r_{3}$, and $r_{4}$) would lead to the
  occurrence of such 
  oppositely directed (Fig.~\ref{img_complexcircuit}) 
  currents \mbox{$i_5 = i_6 \neq 0$} interconnecting both
  parts of the bipolar active region.
  The electrons building up these two currents can be understood
  as the source of the hard \mbox{X-ray} footpoint sources, 
  which are usually 
  observed by \textit{RHESSI} (see Fig.~\ref{img_rhessi}).
  The hard \mbox{X-ray} emission is considered to be 
  generated by energetic electrons via bremsstrahlung
  \citep{1971SoPh...18..489B,1972SoPh...26..441B}. 
  \begin{figure}[t]
     \begin{center}
      \hfill
       \subfigure[The absolute value of the Dreicer field \mbox{$\big|E_\mytext{D}\big|$} is shown
                                  in dependence on the plasma temperature $T$
                                  and electron density \mbox{$N_\mytext{e}$}
                                  for the case \mbox{$\beta_\mytext{D}=\beta_\therm$}.]{\label{img_dreicerfield3d}\includegraphics[width=0.48\textwidth]{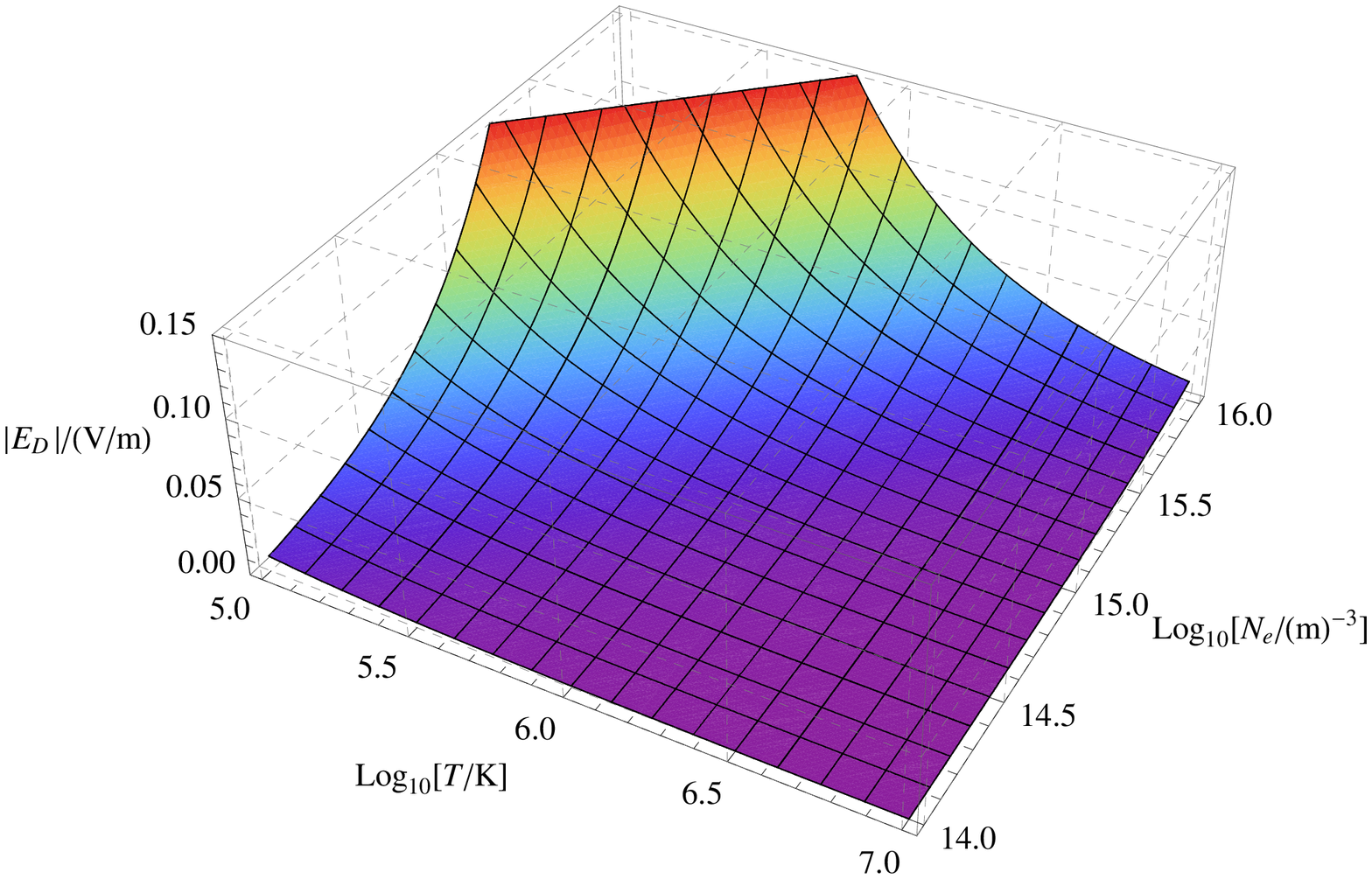}}
       \hfill
       \subfigure[The electric resistivity \mbox{$\eta$} is plotted in dependence
                                      on the temperature for two different electron densities.
                                      The dots in the diagram mark the conditions for
                                      the photosphere (\mbox{$N_\mytext{e}=\unit[4\times 10^{19}]{m^{-3}}$})
                                      and the corona 
                                      (\mbox{$N_\mytext{e}=N_\mytext{co}=\unit[10^{15}]{m^{-3}}$}).]{\label{img_resistivity}\includegraphics[width=0.48\textwidth]{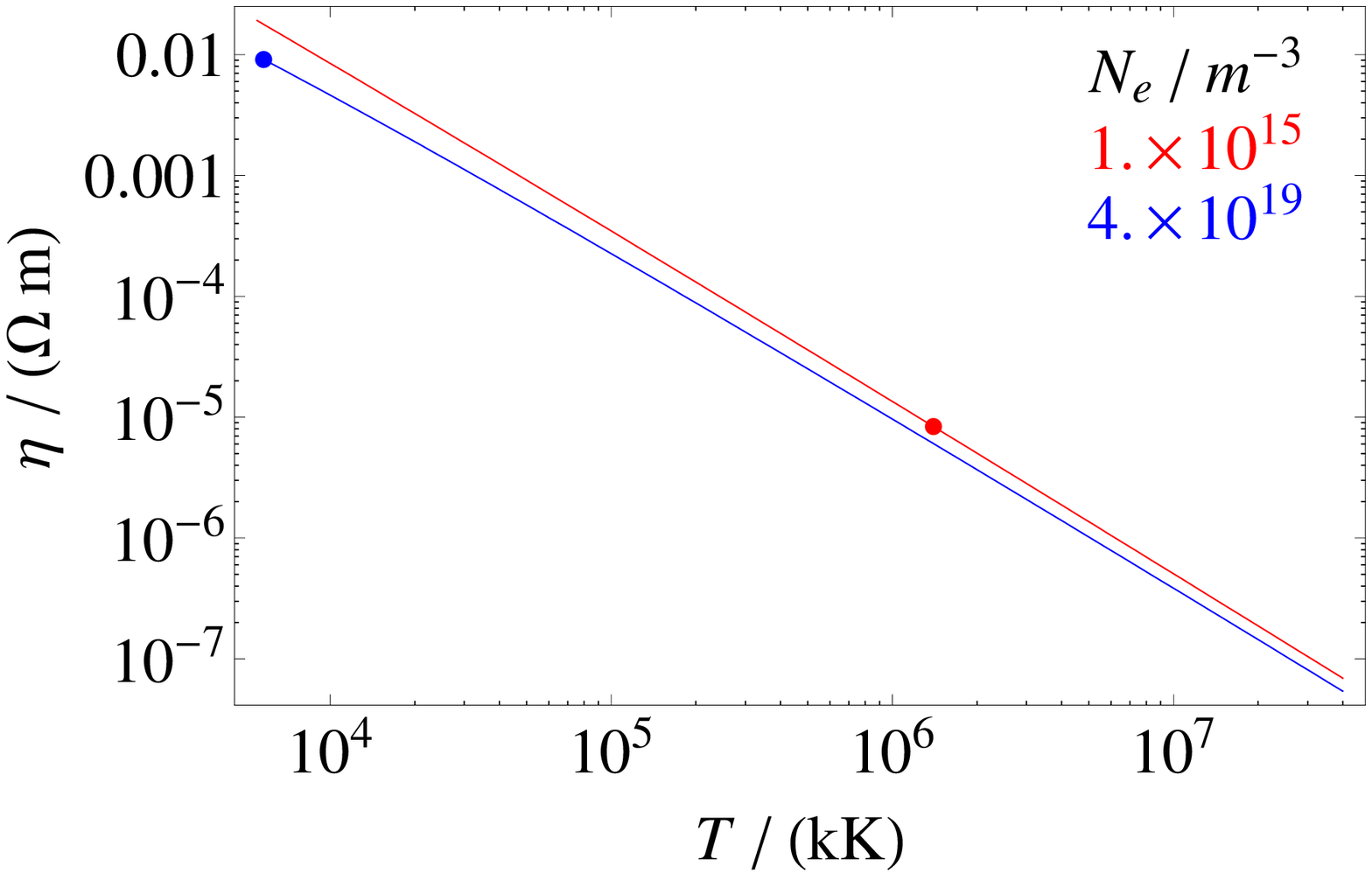}}
       \hfill
       \caption[Dreicer field and Spitzer resistivity]{Dreicer field and Spitzer resistivity both according to \citet{2008PhDT.........9O}}
       \label{img_dreicerfieldandspitzerresistivity}
     \end{center}
    \end{figure}

  Since electric currents always choose the path of the lowest resistance, which is
  through the corona in the considered case, the complete electric circuit
  can be simplified to a single mesh as drawn in Fig.~\ref{img_circuit_secondmesh}.
  In that circuit a DC~power supply $U$ is serially connected with 
  four macroscopic resistors 
  (\mbox{$R_\mytext{ph,$1$}$}, \mbox{$R_\mytext{ph,$2$}$}, 
   \mbox{$r_\mytext{i,$1$}$}, and \mbox{$r_\mytext{i,$2$}$}).
  The two power supplies from Fig.~\ref{img_complexcircuit} are considered to be 
  merged into the one power supply of Fig.~\ref{img_circuit_secondmesh},
  \mbox{i.e.}, \mbox{$U=u_{1} - u_{2}$}.
  Since the coronal contributions
  of the interconnection
  resistors \mbox{$r_{\mytext{i},1}$} and \mbox{$r_{\mytext{i},2}$} 
  from Fig.~\ref{img_complexcircuit} are much smaller than
  the photospheric ones,
  \mbox{$r_{\mytext{i},1}\approx R_\mytext{co,$1$}$}
  and 
  \mbox{$r_{\mytext{i},2}\approx R_\mytext{co,$2$}$}
  is obtained.
  For simplicity 
  \mbox{$R_\mytext{ph}=R_\mytext{ph,$1$}=R_\mytext{ph,$2$}$} and \mbox{$R_\mytext{co}=R_\mytext{co,$1$}=R_\mytext{co,$2$}$}
  are chosen.

  A macroscopic resistor \mbox{$R = \nicefrac{\left(\eta L\right)}{A}$}
  is given by 
  its cross sectional area $A$, length $L$, and 
  electric resistivity $\eta$.
  The first two parameters ($A$ and $L$) are questions of geometry,
  whereas the electric resistivity is a plasma parameter 
  depending strongly on the plasma temperature $T$ and weakly on 
  the electron 
  density \mbox{$N_\mytext{e}$} (see Fig.~\ref{img_resistivity}).
  To choose these parameters, the example shown in Fig.~\ref{img_rhessi}
  is employed: For instance the hard \mbox{X-ray} \mbox{source's} diameter \mbox{$\oslash_\mytext{s}$}
  is assumed to be about 
  \mbox{$\oslash_\mytext{s}\approx\unit[10\times 10^{6}]{m}$}, whereas 
  its depth \mbox{$d_\mytext{s}$} is (according to the depth of
  the photosphere) considered to 
  be \mbox{$d_\mytext{s}\approx\unit[500]{km}$} 
  \citep[see e.g.,][]{book_priest_001}, thus
  \mbox{$A_\mytext{ph} = d_\mytext{s} \oslash_\mytext{s} = \unit[5 \times 10^{12}]{m^{2}}$}
  is obtained.
  \mbox{$L_\mytext{ph}=\unit[40 \times 10^{6}]{m}$} is used for the length
  of the photospheric resistor and the distance between the two hard X-ray sources
  is about \mbox{$L_{\mytext{s}}=\unit[70 \times 10^{6}]{m}$} (see Fig.~\ref{img_rhessi}).
  Assuming the
  overlying magnetic loop to be a semicircle, the length 
  \mbox{$L_\mytext{co}= \nicefrac{\left(\pi L_{\mytext{s}}\right)}{2} \approx \unit[110 \times 10^{6}]{m}$}
  is found.
  The cross sectional area of the loop can be estimated 
  by \mbox{$A_\mytext{s}$} which according to Sec.~\ref{sec_introduction} 
  can be obtained by \textit{RHESSI} observations, \mbox{i.e.},
  \mbox{$A_\mytext{s}\approx \unit[80 \times 10^{12}]{m^2} = A_\mytext{co}$}.
  According to  
  Fig~\ref{img_resistivity} the electric resistivity 
  in the photosphere and corona
  for these given parameters is
  \mbox{$\eta_\mytext{ph}=\unit[9.13 \times 10^{-3}]{\Omega m}$}
  and 
  \mbox{$\eta_\mytext{co}=\unit[8.37 \times 10^{-6}]{\Omega m}$}, respectively.
  With these parameters the values  
  \begin{eqnarray}
     R_\mytext{ph}&=&\frac{\eta_\mytext{ph} L_\mytext{ph}}{A_\mytext{ph}} = \unit[73.0 \times 10^{-9}]{\Omega} \label{eqn_resistor_ph}\\
     R_\mytext{co}&=&\frac{\eta_\mytext{co} L_\mytext{co}}{A_\mytext{co}} = \unit[11.5 \times 10^{-12}]{\Omega} \label{eqn_resistor_co}
   \end{eqnarray} 
  are obtained for the resistors of the circuit shown in Fig.~\ref{img_circuit_secondmesh}.

  The induced voltage 
  \begin{eqnarray}
     U  &=& u_\mytext{ind} B L_\mytext{ph} \label{eqn_inductionpotential1}
  \end{eqnarray}
  is determined using \mbox{Faraday's} induction law,
  if \mbox{$u_\mytext{ind}$} and $B$ denote the speed of the photospheric flow and
  the local magnetic flux density, respectively.
  \mbox{Kirchhoff's} law provides  
  \begin{eqnarray}
     I_{\mytext{mesh}\,\mytext{(II)}} &=& \frac{U}{2\left(R_\mytext{ph}+R_\mytext{co}\right)} \approx \frac{U}{2 R_\mytext{ph}} = u_\mytext{ind} B \, \frac{A_\mytext{ph}}{\eta_\mytext{ph}}  \label{eqn_current1}
  \end{eqnarray}
  for the current (see Fig.~\ref{img_circuit_secondmesh}). 
  Note that the value of the current is independent of the
  length of the photospheric resistor. 
  If it is assumed that the flare power \mbox{$P_\mytext{e}=\unit[10^{22}]{W}$} 
  (see Sec.~\ref{sec_introduction})
  is equal to the electric power in the coronal resistors of the circuit 
  in Fig.~\ref{img_circuit_secondmesh}, then
  \begin{eqnarray}
     I_{\mytext{mesh}\,\mytext{(II)}}  &=& \sqrt{\frac{P_\mytext{e}}{2 R_\mytext{co}}} \approx \unit[2.08\times 10^{16}]{A} \label{eqn_current2}
  \end{eqnarray}
  can be derived by using \mbox{$U_\mytext{co}=R_\mytext{co} I_{\mytext{mesh}\,\mytext{(II)}}$}, 
  where \mbox{$U_\mytext{co}$} represents the potential drop in one of the coronal resistors.
  Since there are two resistors in the corona, 
  the $2$ appears in the denominator of the \mbox{middle-part} of Eq.~(\ref{eqn_current2})
  and in the following denominator of the electric current estimation.
  Note that the current is in good agreement with the electric current
  of about 
  \mbox{$\nicefrac{\left(F_\mytext{e} e\right)}{2} \approx \unit[8 \times 10^{16}]{A}$}, 
  which is generated
  by the observed energetic (\mbox{$\geq \unit[20]{keV}$}) electron 
  flux of about \mbox{$F_\mytext{e}\approx\unitfrac[10^{36}]{1}{s}$} 
  (see Sec.~\ref{sec_introduction}). The quantity $e$ stands for the
  elementary charge.
  By using Eqs.~(\ref{eqn_resistor_ph}), (\ref{eqn_inductionpotential1}),
  (\ref{eqn_current1}), and~(\ref{eqn_current2}) 
  a constraint for the product \mbox{$u_\mytext{ind} B$} can be found
  \begin{eqnarray}
     u_\mytext{ind} B &=& \frac{I_{\mytext{mesh}\,\mytext{(II)}}}{\nicefrac{A_\mytext{ph}}{\left(2\eta_\mytext{ph}\right)}} \approx \unitfrac[76.1]{V}{m}. \label{eqn_inductionparameters1}
  \end{eqnarray}
  This requirement (Eq.~(\ref{eqn_inductionparameters1})) can be fulfilled,
  \mbox{e.g.}, for 
  \mbox{$u_\mytext{ind}\approx\unitfrac[870]{m}{s}$} and
  \mbox{$B\approx\unit[0.087]{T}$}, 
  which are reasonable conditions for the photosphere \citep[see e.g.,][]{2004ApJ...617L.151Y}.
  Finally the electric field along the coronal loop is estimated to be 
     \mbox{$E = -\nicefrac{U_\mytext{co}}{L_\mytext{co}}\approx\unitfrac[-2.18 \times 10^{-3}]{V}{m}$}.
  The potential drop at one of the coronal resistors is \mbox{$U_\mytext{co} \approx \unit[240]{kV}$},
  and Eq.~(\ref{eqn_inductionpotential1}) gives 
  \mbox{$U\approx\unit[3\times 10^9]{V}$}
  for the power \mbox{supply's} voltage.

  In order to asses whether the photospheric flow can support 
  the flare the following estimation is done:
  The continuous photospheric plasma motion along the way
  \mbox{$\dd{X}=u_\mytext{ind} \dd{t}$} in the time 
  interval \mbox{$\dd{t}$} builds up an energy of
  \begin{eqnarray}
   \dd{W_\mytext{s}}&=&(N_\mytext{ph,e} A_\mytext{ph} L_\mytext{ph}) \, (e u_\mytext{ind}\, B) \, \dd{X}, \label{eqn_powerestimation1}
  \end{eqnarray}
  due to the action of the 
  Lorentz force \mbox{$e u_\mytext{ind}\, B$}.
  Here 
  \mbox{$N_\mytext{ph,e}$}, and 
  \mbox{$N_\mytext{ph,e} A_\mytext{ph} L_\mytext{ph}$}  
  stand for 
  the total electron number density in the photosphere, and
  the total number of electrons in the volume of \mbox{$A_\mytext{ph}\times L_\mytext{ph}$},
  respectively.
  Therefore the power of the photospheric motion
  \begin{eqnarray}
   P_\mytext{ph}=\ddfrac{W_\mytext{s}}{t}&=&(N_\mytext{ph,e} A_\mytext{ph} L_\mytext{ph}) \, (e u_\mytext{ind}\, B) \, u_\mytext{ind} \nonumber\\ 
                 &=& e N_\mytext{ph,e} A_\mytext{ph} L_\mytext{ph}\, B\, u_\mytext{ind}^2 \approx \unit[8.4\times 10^{25}]{W}  \label{eqn_powerestimation2}
  \end{eqnarray}
  (when
  \mbox{$L_\mytext{ph}=\unit[4\times 10^{7}]{m}$},
  \mbox{$u_\mytext{ind}=\nicefrac{\dd{X}}{\dd{t}}=\unitfrac[870]{m}{s}$}, and
  \mbox{$B=\unit[0.087]{T}$},
  as introduced in Sec.~\ref{sec_complexcircuit}, and
  \mbox{$N_\mytext{ph,e}=\unit[4\times 10^{19}]{m^{-3}}$}
  are used) is much higher 
  than the required flare released power in the corona. 
  Hence the photospheric motion possesses more than 
  enough power needed for the flare.

\typeout{=====================================================}
\typeout{Electron Acceleration}
\section{Electron Acceleration}\label{sec_electronacceleration}
  \begin{figure}[t]
     \begin{center}
      \includegraphics[width=0.7\textwidth]{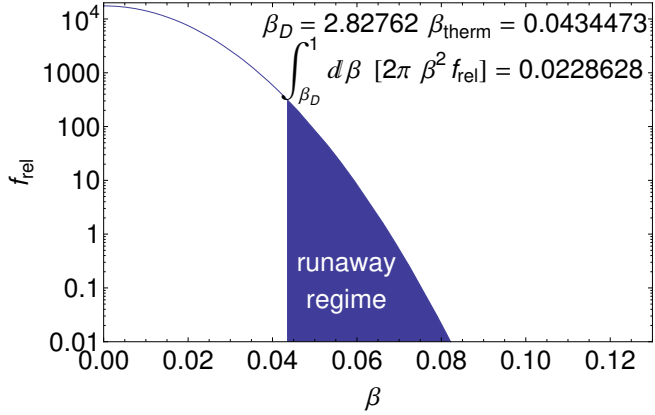}
       \caption[Maxwellian velocity distribution function]{A to unity normalised one dimensional relativistic Maxwellian velocity distribution \mbox{$f_\mytext{rel}$} plotted for a temperature of \mbox{$T=\unit[1.4]{MK}$} is presented as a function of the normalised electron velocity $\beta$. The runaway region \mbox{$\beta_\mytext{D}<\beta$} is filled, \mbox{i.e.}, about \mbox{$2.3\%$} of the total electrons are accelerated when an electric field of \mbox{$E\approx\unitfrac[-2.18 \times 10^{-3}]{V}{m}$} is chosen.}
       \label{img_maxwelliandistributionrel}
     \end{center}
  \end{figure}
   Electrons gain energy only by electric field acceleration.
   In this paper the electron acceleration is caused by the electric field
   related to the potential drop at the coronal resistor.
   Since the electric field is aligned along the magnetic field in
   the coronal loop, the equation of motion for an 
   electron\footnote{The electron is supposed to carry the negative elementary charge \mbox{$- e$} and to possess the rest mass \mbox{$m_\mytext{e}$}.}
   can be written in its \mbox{one-dimensional} form
   \begin{eqnarray}
      \ddfrac{p}{t} & = &  -e E - m_\mytext{e}\, \mytext{sign}[p] \, \big| D \big|. \label{eqn_lawofmotion01}
   \end{eqnarray}
   The quantities \mbox{$t$} and 
   \mbox{$p=\nicefrac{\left(m_\mytext{e} c\,\beta\right)}{ \sqrt{1-\beta^{2}} }$}
   denote the time and the \mbox{electron's} momentum, 
   respectively.
   The normalised electron velocity $\beta$ is given 
   by \mbox{$\beta = \nicefrac{V}{c}$} 
   with $V$ as the \mbox{electron's} velocity. The quantity 
   $D$ stands for the
   \mbox{electron's} deceleration due to Coulomb collisions 
   \citep{2007A&A...463.1143O}.
   By inserting the expression for the \mbox{electron's} momentum into
   Eq.~(\ref{eqn_lawofmotion01})
   \begin{eqnarray}
     \ddfrac{p}{t} & = & m_\mytext{e} c \gamma^3 \ddfrac{\beta}{t} = -e E - m_\mytext{e}\, \mytext{sign}[\beta] \, \big| D \big|  \label{eqn_lawofmotion02}
   \end{eqnarray}
   is obtained, if 
   \mbox{$\gamma = \nicefrac{1}{\sqrt{1-\beta^{2}}}$} for the Lorentz factor \mbox{$\gamma$}
   is used.
   The momentum change per arc length \mbox{$x$} along the magnetic field
   can be derived from Eq.~(\ref{eqn_lawofmotion02}), \mbox{i.e.},
   \mbox{$\ddnicefrac{p}{x} = m_\mytext{e} c \, \gamma^3 \cdot \ddnicefrac{\beta}{x} = \nicefrac{1}{\beta c} \cdot \ddnicefrac{p}{t}$}.

   An electron travelling through an
   \mbox{electron-proton} plasma experiences Coulomb collisions,
   which always make the electron lose energy. Thus the electron
   momentum changes by deceleration, as described in the very last
   sum of Eq.~(\ref{eqn_lawofmotion02}).
   The Coulomb deceleration $D$ in an \mbox{electron-proton}
   plasma has two contributions, namely  
   the \mbox{electron-electron} interaction \mbox{$D_\mytext{e}$} and
   the \mbox{electron-proton} interaction \mbox{$D_\mytext{p}$},
   \mbox{i.e.}, \mbox{$D=D_\mytext{e}+D_\mytext{p}$}.
   Each of these contributions is given by
   \begin{eqnarray}
     \forall l \in \{\mytext{e},\mytext{p}\}: \, D_l &=&  \frac{Z_l^2 e^4 \, N_l     \ln[\Lambda_l]}{4 \pi \epsilon_0^2 \left(\nicefrac{1}{m_\mytext{e}}+\nicefrac{1}{m_l}\right)^{-2} c^2 \beta_l^2} \quad\mytext{for \mbox{$\!\quad\beta_l \not= 0$}}\label{eqn_deceleration}
   \end{eqnarray}
   \citep[see e.g.,][]{2007A&A...463.1143O}.
   The quantity \mbox{$\epsilon_0$} stands for the permittivity of free space, 
   whereas
   \mbox{$Z_l$} represents the 
   charge number\footnote{In a fully ionised \mbox{electron-proton} plasma \mbox{$Z_\mytext{e}=1$} and \mbox{$Z_\mytext{p}=1$} is satisfied.},
   \mbox{$m_l$} the rest mass, and
   \mbox{$N_l$} the number density of the
   particle species
   $l\in \{ \mytext{e}$~for ``electron''$, \mytext{p}$~for ``proton''$\}$.
   \mbox{$\beta_l$} represents the relative velocity of the electron 
   with respect to the electrons and protons of the plasma, in which the
   electron propagates. Hence
   \mbox{$\beta_\mytext{e}=\sqrt{\beta^2+3\beta_\therm^2}$} and  
   \mbox{$\beta_\mytext{p}\approx\beta$}
   are the relative velocities of a moving electron with respect to electrons 
   and protons of the (background) plasma.
   \mbox{$\beta_\therm =  \sqrt{\nicefrac{\left(k_\mytext{B} T\right)}{\left(m_\mytext{e} c^2\right)}} $} 
   is the thermal velocity normalised to the speed of light.
   The Coulomb logarithm 
    \mbox{$\ln\left[\Lambda_{j}\right]=\ln\left|\sqrt{ \nicefrac{ \left(\lambda_\mytext{D}^2 + b_{0,j}^2 \right)}{ \left(2 b_{0,j}^2 \right)} }\right|$}
   contains the Debye length 
    \mbox{$\lambda_\mytext{D}=\sqrt{\nicefrac{\left(\epsilon_0 k_\mytext{B} T\right)}{\left(N_\mytext{e} e^2\right)}}$}
   and the Coulomb collision impact parameter 
    \mbox{$ b_{0,j} = \left|\frac{e^2}{4\pi \epsilon_0 c^2}    \frac{Z_j}{\left(\nicefrac{1}{m_\mytext{e}}+\nicefrac{1}{m_j}\right)^{-1} \beta_j^2} \right|$}.

   As it can be seen from Eq.~(\ref{eqn_lawofmotion02}) a special electric
   field, the so called \cite{1959PhRv..115..238D,1960PhRv..117..329D} field
   \begin{eqnarray}
     E_\mytext{D} =  - \left( \frac{ m_\mytext{e}}{e} \,\mytext{sign}[\beta] \big| D\big| \right)\Bigg|_{\beta=\beta_\mytext{D}}     \label{eqn_dreicerfield01}
   \end{eqnarray}
   exists, which makes the time derivative of the \mbox{electron's} momentum 
   vanish by definition, 
   \mbox{i.e.}, \mbox{$\nicefrac{\mytext{d}p}{\mytext{d}t}=0$}.
   Thus there is a critical velocity, the \mbox{so-called} Dreicer velocity $\beta_\mytext{D}$,
   which characterises the change of the \mbox{electron's} momentum:
   If the initial electron velocity \mbox{$\big|\beta_\mytext{i}\big|$} 
   is greater than the Dreicer velocity
   (\mbox{$\big|\beta_\mytext{i}\big|>\big|\beta_\mytext{D}\big|$}), the electrons
   are accelerated nearly collisionless and are called 
   runaway electrons \citep[e.g.,][]{1995ApJ...452..451H}.
   In the opposite case  
   (\mbox{$\big|\beta_\mytext{i}\big|<\big|\beta_\mytext{D}\big|$})
   they are thermalised due to the strong Coulomb collisions.
   The Fig.~\ref{img_dreicerfield3d} shows the
   dependence of the absolute value of \mbox{$E_\mytext{D}$} 
   on \mbox{$N_\mytext{e}$} and \mbox{$T$} 
   for \mbox{$\beta_\mytext{D}=\beta_\therm$}.

   The previously determined electric field of 
   \mbox{$E\approx\unitfrac[-2.18 \times 10^{-3}]{V}{m}$} in the corona
   corresponds to \mbox{$\beta_\mytext{D} \approx 2.83\,\beta_\therm$} and 
   therefore all the electrons which possess a normalised velocity 
   higher than \mbox{$2.83\,\beta_\therm$} are located in the runaway regime.
   If an one dimensional relativistic Maxwellian velocity distribution 
   at a temperature of $\unit[1.4]{MK}$ is assumed then 
   about \mbox{$2.3\%$} of the total available electrons are frictionless 
   accelerated by this electric field.

 \typeout{=====================================================}
 \typeout{Conclusions}
 \section{Conclusions}\label{sec_conclusions}
   Electron acceleration is a very important process during solar
   flares, since a large fraction of the released flare
   energy is deposited in the energetic electrons
   \citep{1971SoPh...17..412L,2004JGRA..10910104E}.
   Such electrons can be observed by \mbox{in-situ} 
   spacecraft measurements, 
   by their \mbox{non-thermal} radio signatures, and by 
   the \mbox{X-ray} 
   radiation which they emit, if they travel through the 
   dense chromosphere.
   The paper at hand explains how a large scale DC~electric
   field is generated in the corona, which subsequently can accelerate 
   the electrons:
   The photospheric motion in bipolar active regions can generate 
   electric powers high enough to drive a system of electric circuits 
   in the solar atmosphere.
   The photospheric motion can induce voltages, because
   the photospheric plasma is not fully ionised, \mbox{i.e.}, 
   it contains a substantial part of neutral particles.
   Due to the low resistivity of the coronal plasma in comparison
   to the photospheric one, the system of electric circuits is 
   closed through the corona,
   if there is a magnetic connection between the two regions of 
   the bipolar active region. Then a large scale DC~electric 
   field can establish in the corona which can be held responsible
   for the electron acceleration.

   Adopting plausible values
   of parameters in the solar atmosphere (as presented in this paper),
   an electric field of about \mbox{$\unitfrac[-2]{mV}{m}$} is generated 
   in the corona. Such a field can accelerate sufficient (depending on
   the plasma temperature more than \unit[2]{\%} of the total) electrons 
   to kinetic energies of up to 
   around \unit[240]{keV} in typical time scales of 
   less than a second \citep{2002SSRv..101....1A}.
   The resulting fluxes of energetic electrons are of the order of
   \mbox{$\unitfrac[10^{36}]{\mytext{1}}{s}$},
   as it is really observed with the 
   \textit{RHESSI} spacecraft \citep{2007CEAB...31..135W}. 
   
   According to \mbox{Amp\`{e}re's} law electric currents are
   the sources for magnetic fields. Basing on
   \mbox{$F_\mytext{e}\approx\unitfrac[10^{36}]{1}{s}$} and considering
   (as usually observed) two hard \mbox{X-ray} sources,
   the electrical current related with the \mbox{X-ray} emission can be
   estimated by
    \mbox{$I = \nicefrac{e F_\mytext{e}}{2} = \unit[8 \times 10^{16}]{A}$}.
   Such a current localised in a
   coronal loop would be related to a magnetic flux density
   in the order of 
    \mbox{$B \approx \nicefrac{\mu_{0} I}{L_{\mytext{s}}} \approx \unit[1.44]{kT}$},
   where \mbox{$\mu_{0}$} represents the vacuum permeability.
   On one hand these extremely high magnetic fluxes are not 
   observed in the solar atmosphere, on the other hand the
   fluxes of energetic electrons deduced from the 
   hard \mbox{X-ray} observations (\mbox{e.g.}, \textit{RHESSI})
   are related with such high electric currents of the order of
   \mbox{$\unit[10^{17}]{A}$}.
   In order to cancel the high magnetic fields induced
   in the corona by these high currents, the coronal 
   currents need to be 
   oppositely directed and cospatially located 
   to each other.
   The system
   of electric circuits as discussed here (see Fig.~\ref{img_complexcircuit}) 
   has two electric currents (\mbox{i.e.}, \mbox{$i_{5}$} and \mbox{$i_{6}$}) 
   through the corona. Note that the induced current is
   independent from \mbox{$L_\mytext{ph}$}, \mbox{i.e.} the photospheric length scale  
   (see Eq.~(\ref{eqn_current1})).
   
   The important conclusions of the model presented in this paper
   are:
   \begin{enumerate}
     \item The energy generated by the photospheric plasma motion 
           is transported electrically into the corona where it
           is transferred into the flare. 
     \item The photospheric plasma motion has enough power to support
           a flare in the corona. 
     \item The electric fields related with the potential drops at the coronal
           resistors are oppositely directed, so that the electrons are accelerated
           to two opposite directions leading to a double source of the hard
           \mbox{X-ray} radiation, as it is usually observed.
   \end{enumerate}
   The hard \mbox{X-ray} radiation is produced by bremsstrahlung
   of the energetic electrons in the dense chromospheric plasma
   \citep{1971SoPh...18..489B,1972SoPh...26..441B}.      

   In summary, in the solar atmosphere electric circuits can be driven by an electric power 
   supply induced by the photospheric motion in a bipolar active region.
   If these electric circuits are closed via the corona, an electric 
   field is established along a coronal loop. Electrons can be accelerated
   along these large scale DC~electric fields. The so produced energetic 
   electrons can emit hard \mbox{X-ray} radiation. 
   Therefore the model presented is in good agreement with observations. 

 \typeout{=====================================================}
 \typeout{Acknowledgements}
 \section*{Acknowledgements} \label{sec_acknowledgements}
  This work was supported by \textit{Deutsches Zentrum f\"u{}r \mbox{Luft-} und Raumfahrt}
  under DLR grant \mbox{50QL0001}.

 \typeout{=====================================================}
 \typeout{References (BiBTeX)}
 \bibliographystyle{ceab}
 \bibliography{onel}

%
 \typeout{********************************************************}
 \typeout{ End of document}
 \typeout{********************************************************}
\end{document}